\documentclass[
  ]
  {aipproc}

\layoutstyle{6x9}

\usepackage{fancyhdr}

\begin{document}

\title{Extremely long hard bursts observed by Konus-Wind}

\author{V. Pal'shin}{
  address={Ioffe Physical-Technical Institute of the Russian
Academy of Sciences, St. Petersburg, 194021, Russia} }
\author{R. Aptekar}{
  address={Ioffe Physical-Technical Institute of the Russian
Academy of Sciences, St. Petersburg, 194021, Russia} }
\author{D. Frederiks}{
  address={Ioffe Physical-Technical Institute of the Russian
Academy of Sciences, St. Petersburg, 194021, Russia} }
\author{S. Golenetskii}{
  address={Ioffe Physical-Technical Institute of the Russian
Academy of Sciences, St. Petersburg, 194021, Russia} }
\author{V. Il'inskii}{
  address={Ioffe Physical-Technical Institute of the Russian
Academy of Sciences, St. Petersburg, 194021, Russia} }
\author{E. Mazets}{
  address={Ioffe Physical-Technical Institute of the Russian
Academy of Sciences, St. Petersburg, 194021, Russia} }
\author{K. Yamaoka}{
  address={Department of Physics and Mathematics, Aoyama Gakuin University,
5-10-1, Fuchinobe, Sagamihara 229-8558} }
\author{M. Ohno}{
  address={Department of Physical Sciences, School of Science, Hiroshima University
1-3-1 Kagamiyama, Higashi-Hiroshima, Hiroshima 739-8526} }
\author{K. Hurley}{
  address={Space Sciences Laboratory, University of California
at Berkeley, Berkeley, CA 94720-7450, USA} }
\author{T. Sakamoto}{
  address={Goddard Space Flight Center, NASA, Greenbelt, MD
20771, USA} }
\author{P. Oleynik}{
  address={Ioffe Physical-Technical Institute of the Russian
Academy of Sciences, St. Petersburg, 194021, Russia} }
\author{M. Ulanov}{
  address={Ioffe Physical-Technical Institute of the Russian
Academy of Sciences, St. Petersburg, 194021, Russia} }
\author{I. G. Mitrofanov}{
  address={Institute for Space Research, Profsojuznaja 84/32, Moscow 117997, Russia} }
\author{D. Golovin}{
  address={Institute for Space Research, Profsojuznaja 84/32, Moscow 117997, Russia} }
\author{M. L. Litvak}{
  address={Institute for Space Research, Profsojuznaja 84/32, Moscow 117997, Russia} }
\author{A. B. Sanin}{
  address={Institute for Space Research, Profsojuznaja 84/32, Moscow 117997, Russia} }
\author{W. Boynton}{
  address={Lunar and Planetary Laboratory, University of Arizona, Tucson, AZ 85721, USA} }
\author{C. Fellows}{
  address={Lunar and Planetary Laboratory, University of Arizona, Tucson, AZ 85721, USA} }
\author{K. Harshman}{
  address={Lunar and Planetary Laboratory, University of Arizona, Tucson, AZ 85721, USA} }
\author{C. Shinohara}{
  address={Lunar and Planetary Laboratory, University of Arizona, Tucson, AZ 85721, USA} }
\author{R. Starr}{
  address={Department of Physics, The Catholic University of America, Washington, DC , 20064, U.S.A} }


\thispagestyle{fancy}

\cfoot{Copyright (2006) American Institute of Physics. This article
may be downloaded for personal use only. Any other use requires
prior permission of the author and the American Institute of
Physics.}


\rhead{The following article appeared in AIP Conf. Proc. 1000, pp.
117-120 (2008) and may be found at
http://link.aip.org/link/?apc/1000/117.}

\begin{abstract}
We report the observations of the prompt emission of the extremely
long hard burst, GRB 060814B, discovered by Konus-Wind and localized
by the IPN. The observations reveal a smooth, hard, $\sim$40-min
long pulse followed by weaker emission seen several hours after the
burst onset. We also present the Konus-Wind data on similar burst,
GRB 971208, localized by BATSE/IPN. And finally we discuss the
different possible origins of these unusual events.
\end{abstract}

\maketitle

\thispagestyle{fancy}

\section{Introduction}
Konus-Wind is a gamma-ray all-sky spectrometer~\citep{Aptekar1995}
which has been successfully operating since November 1994. Wind
orbit is far from the Earth magnetosphere (at distance of $\sim$1--7
light seconds) that enables nearly uninterrupted observations of all
sky under very stable background. In the waiting mode Konus-Wind
measures the count rates in three energy bands which covers the
$\sim$15--1000 keV range with accumulation time of 2.944 s. This
mode enables observations of ultra long gamma-ray bursts with
duration $\geq$500~s. During 13 years of observations Konus-Wind has
detected two extremely long, single pulsed, hard GRBs.

\section{GRB 060814B}
\begin{figure}[t!]
\includegraphics[width=0.75\textwidth]{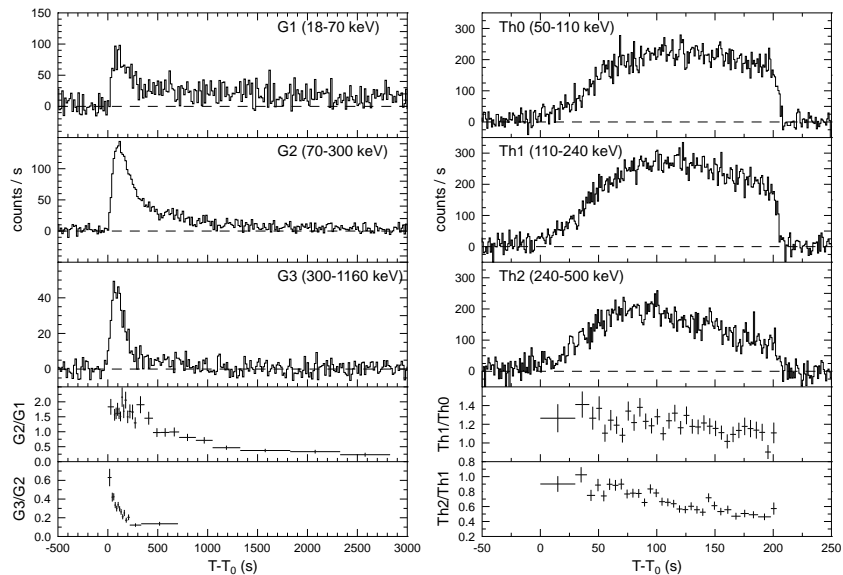}
\caption{Konus-Wind (\textit{Left}) and Suzaku-WAM (\textit{Right})
background subtracted light curves of GRB 060814B in three energy
bands and the hardness ratios, T$_0$=37070 s UT (10:17:50).}
\label{lc060814B}
\end{figure}

A very long burst was detected by Konus-Wind in the waiting mode at
2006-08-14 T$_0$=37070 s UT (10:17:50). The burst light curve shows
a single, smooth, FRED-like pulse with a duration of $\simeq$2700~s.
The most intense part of the burst (initial several hundreds
seconds) was also detected by Ulysses, Mars Odyssey (HEND),
Suzaku-WAM, and INTEGRAL-SPI-ACS. That enabled to localize the burst
to the $\simeq$7.1~deg$^2$ 3$\sigma$ IPN error box, whose
coordinates are given in Table~\ref{TableBox}. The Suzaku-WAM light
curve shows a sharp decline at T-T$_0 \simeq$200~s due to source set
below the horizon (see Fig.~\ref{lc060814B}). This occultation step
let us to derive much smaller combined IPN/WAM box with the area of
$\simeq$0.6~deg$^2$ (given in Table~\ref{TableBox}). The galactic
coordinates of the box center are $l$, $b$  = 162.88, +6.06 deg.

The Konus-Wind light curve of GRB~060814B in three energy bands is
shown in Fig.~\ref{lc060814B}. The burst demonstrates strong
hard-to-soft spectral evolution with the hardest spectrum at the
burst onset. After the end of the main pulse there is an extended
emission seen in the G1 and G2 bands during several hours (4 hours
in the G2 at $>4\sigma$-level and even longer in the G1). The
emission was detected by the same detector which observed the main
pulse, so it might be a burst tail. Unfortunately the localization
area occurred in the BAT FoV only in $\simeq$11~hours after the
burst onset. Nothing evident was found in the BAT data of the 500-s
long observation (obsID 00035631001).

The Suzaku-WAM multichannel spectrum accumulated from T$_0$ to
T$_0$+200~s is well fitted (in the 80--2000 keV range) by
CPL model: $dN/dE \propto E^{-\alpha} \exp[-(2-\alpha)E/E_p]$ with
$\alpha = 0.19 \pm 0.14$, $E_p =477(-27,+31)$~keV ($\chi^2$=21/23
dof). The errors are given at 90\% confidence level. The Konus-Wind
3-channel spectrum for the same interval yields $\alpha=0.89 \pm
0.07$, $E_p=544 \pm 52$~keV. The errors are estimated at 1$\sigma$
level by propagation of errors in the observed counts (here and
below). Using the KW 3-channel data we also estimated the spectral
parameters for whole burst, for the hard pulse, and for the possible
tail. The results are given in Table~\ref{TableSpParam}. The
estimated burst fluence is $(2.35 \pm 0.22) \times
10^{-4}$~erg~cm$^{-2}$ (18--1170 keV). The fluence of the tail is
$(2.1 \pm 0.2) \times 10^{-4}$~erg~cm$^{-2}$ in the same range.

\thispagestyle{fancy}
\cfoot{Copyright (2006) American Institute of Physics. This article
may be downloaded for personal use only. Any other use requires
prior permission of the author and the American Institute of
Physics.}

\section{GRB 971208}
\begin{figure}[t!]
\includegraphics[width=0.75\textwidth]{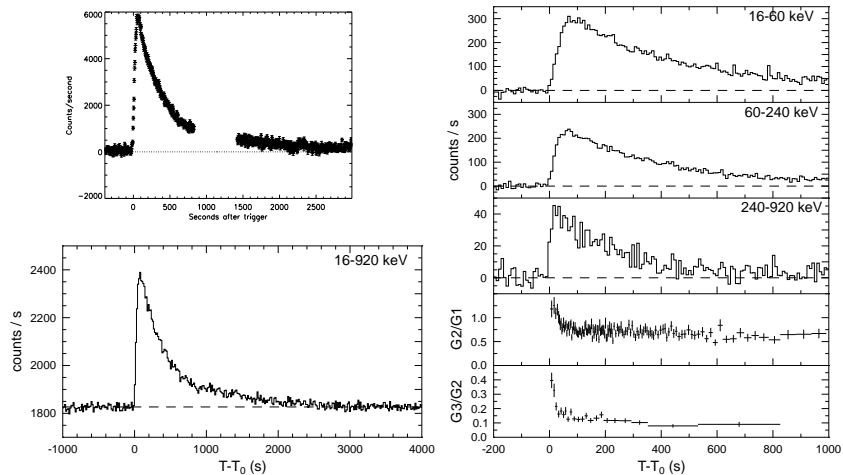}
\caption{\textbf{Left}: Konus-Wind (\textit{Bottom}) and BATSE light
curves (\textit{Top}: background subtracted, 20--1000 keV range) of
GRB 971208. \textbf{Right}: Konus-Wind background subtracted light
curve in three energy bands and the hardness ratios. T$_0$=28092 s
UT (07:48:12).} \label{lc971208}
\end{figure}

An unusually long, smooth, single pulsed gamma-ray burst was
detected by BATSE at 1997-12-08 28092.1295 s UT (trigger 6526:
~\citep{Connaughton1997}). This burst was observed by Konus-Wind in
full entirety (see Fig.~\ref{lc971208}). The total burst duration is
$\simeq$2500~s. One can see in Fig.~\ref{lc971208} the BATSE light
curve which shows a similar duration. There is no any sign of
extended emission after T-T$_0 \simeq$2500~s.

The burst demonstrates a similar to GRB 060814B hard-to-soft
spectral evolution. Using the KW 3-channel data we estimated the
spectral parameters for the hard pulse, and for the whole burst (see
Table~\ref{TableSpParam}). The estimated burst fluence is $(2.55 \pm
0.11) \times 10^{-4}$~erg~cm$^{-2}$ (15--1000 keV). The fluence
reported by BATSE for the first 800 s is $(1.86 \pm 0.03) \times
10^{-4}$~erg~cm$^{-2}$ in the 25-1800 keV
range~\citep{Connaughton1997}.

\thispagestyle{fancy}

\section{Discussion}
Two obvious factors can make longer the burst: cosmological time
dilation and relativistic curvature effect. Assuming, for example,
$z=10$ we would have $E_{iso} \simeq 2\times 10^{55}$~erg,
$\Delta$T$_{rest} \simeq 230$~s, and $E_{p, rest} \simeq$3700, and
1600 ~keV correspondingly for GRB 060814B and 971207. Such $E_{iso}$
is about a factor of 10 greater than the largest known value. But we
should take in account a high probability of lensing for such
redshift. $E_{p, rest} \simeq$3700~keV seems to be a bit large, but
it is not exceptional: Swift GRB 050717 had $E_p \simeq
1900$~keV~\citep{Krimm2006}, so even assuming $z \simeq 1$, it would
correspond to $E_{p, rest} \simeq$3800 keV. Hence, these bursts
might be high redshift GRBs magnified by lensing.
The pulse shapes and the character of spectral evolution indicate
that the curvature effect
(e.g.~\citep{Kocevski2003},\citep{Qin2006}) may play a main role in
forming the pulses. For pulses created purely by the curvature
effect $\Delta$T$\propto (1+z) R_s/\Gamma^2$. Hence, very long
duration may be due to unusually large shell radius $R_s$. That
requires some specific conditions in the circumburst medium and/or
in the shock.

%
\begin{table}
\begin{tabular}{lccccc}
\hline &\tablehead{2}{c}{t}{IPN box (J2000)}&\hspace{3mm}&\tablehead{2}{c}{t}{Combined IPN/WAM box (J2000)}\\
\hline
  & \tablehead{1}{c}{t}{RA, deg}
  & \tablehead{1}{c}{t}{Dec, deg}
  &
  & \tablehead{1}{c}{t}{RA, deg}
  & \tablehead{1}{c}{t}{Dec, deg}\\
\hline
Center & 81.98 & +47.85 && 81.19 & +46.60\\
Corner1 & 83.20 & +46.76 && 80.73 & +48.66\\
Corner2 & 80.86 & +44.66 && 80.76 & +47.55\\
Corner3 & 80.72 & +48.94 && 81.42 & +45.19\\
Corner4 & 83.30 & +51.11 && 81.66 & +45.40\\
\hline
\end{tabular}
\caption{Localization of GRB 060814B}
\label{TableBox}
\end{table}

\begin{table}
\begin{tabular}{ccccccc}
\hline \tablehead{3}{c}{t}{GRB 060814B}&\hspace{3mm}&\tablehead{3}{c}{t}{GRB 971208}\\
\hline
  \tablehead{1}{c}{b}{Time interval\\s}
  & \tablehead{1}{c}{b}{$\alpha$\\}
  & \tablehead{1}{c}{b}{$E_p$\\keV}
  &
  &  \tablehead{1}{c}{b}{Time interval\\s}
  & \tablehead{1}{c}{b}{$\alpha$\\}
  & \tablehead{1}{c}{b}{$E_p$\\keV}\\
\hline 0--700 & 0.91$\pm$0.07 & 374$\pm$30 &&
0--485 & 1.246$\pm$0.033 & 165$\pm$7\\
0--2700 & 1.46$\pm$0.07 & 341$\pm$61 &&
0--2500 & 1.185$\pm$0.075 & 144$\pm$12\\
2700--12300\tablenote{spectrum was fitted by a simple PL} & 2.7$\pm$0.2 & -- && & &\\
\hline
\end{tabular}
\caption{Spectral parameters of GRB 060814B and GRB 971208}
\label{TableSpParam}
\end{table}

\thispagestyle{fancy}

\cfoot{Copyright (2006) American Institute of Physics. This article
may be downloaded for personal use only. Any other use requires
prior permission of the author and the American Institute of
Physics.}

\begin{theacknowledgments}
We are grateful to Valeri Connaughton for kindly providing us with
the figure of the BATSE light curve of GRB 971208. The Konus-Wind
experiment is supported by a Russian Space Agency contract and RFBR
grant 06-02-16070. V.P. and D.F. thank Neil Gehrels and USRA for the
support of their participation in the conference. KH is grateful for
support under the Mars Odyssey Participating Scientist program (JPL
Contract 1282043), the INTEGRAL U.S. Guest Investigator program
(NASA Grant NNX07AJ65G), and the Suzaku U.S. Guest Investigator
program (NASA Grant NNX06AI36G).
\end{theacknowledgments}

\bibliographystyle{aipproc}
\bibliography{GRB2007}
%
%
%
%
%
%
%
\end{document}